\documentclass[aps,twocolumn,floats,prd,nofootinbib,10pt,longbibliography,superscriptaddress]{revtex4-1}

\usepackage{comment}
\usepackage[dvips]{graphicx} %
\usepackage{graphicx,amsmath,amsfonts,amssymb,slashed,float,hyperref}
\usepackage[normalem]{ulem}
\usepackage{bbold,wasysym}
\usepackage{graphicx}
\usepackage{array,multirow}
\usepackage[utf8]{inputenc}
\usepackage{scalerel}

\usepackage[usenames,dvipsnames]{xcolor} 

\usepackage{soul}
\usepackage{bm}

\definecolor{RoyalBlue}{rgb}{0.25,.41,.88}
\definecolor{celestialblue}{rgb}{0.29, 0.59, 0.82}

\setstcolor{Blue}

\def\AB#1{\textcolor{Magenta}{AB}}

\newcommand{\be}{\begin{equation}}
\newcommand{\ee}{\end{equation}}
\newcommand{\bea}{\begin{eqnarray}}
\newcommand{\eea}{\end{eqnarray}}
\newcommand{\Beq}{\begin{equation}\begin{aligned}}
\newcommand{\Eeq}{\end{aligned}\end{equation}}

\definecolor{cerulean}{rgb}{0., 0.62,0.7}

\newcommand{\mpl}{m_{\rm pl}}

\newcommand{\lcdm}{{\Lambda\textrm{CDM}}}

\newcommand{\dvp}{{\frac{\delta\hspace{-0.05em}\varphi}{\varphi_{{\scaleto{\rm env}{1.8pt}}}}}}

\usepackage{color}
\usepackage{ifthen}
\newboolean{editorial}
\setboolean{editorial}{true}
\newcommand{\editorial}[2]{\ifthenelse{\boolean{editorial}}{\textcolor{red}{[\textsf{\textbf{{#1}}}: }\textcolor{blue}{\textsf{{#2}}}\textcolor{red}{]}}{}}

\usepackage{xcolor}

\begin{document}

\title{A novel integrated Sachs-Wolfe effect from Early Dark Energy}

\author{Tristan~L.~Smith}
\affiliation{Department of Physics and Astronomy, Swarthmore College, Swarthmore, PA 19081, USA}
\affiliation{Center for Cosmology and Particle Physics, Department of Physics,
New York University, New York, NY 10003, USA}

\author{John~T.~Giblin~Jr.}
\affiliation{Department of Physics, Kenyon College, 201 N College Rd, Gambier, OH 43022, USA}
\affiliation{Department of Physics/CERCA/Institute for the Science of Origins, Case Western Reserve University, Cleveland, OH 44106-7079 -- USA}

\author{Mustafa~A.~Amin}
\affiliation{Department of Physics and Astronomy, Rice University, Houston, TX 77005, USA}

\author{Mary~Gerhardinger}
\affiliation{Department of Physics, Kenyon College, 201 N College Rd, Gambier, OH 43022, USA}
\affiliation{Center for Particle Cosmology, Department of Physics and Astronomy,
University of Pennsylvania, Philadelphia, Pennsylvania 19104, USA}

\author{Ericka~Florio}
\affiliation{Department of Physics, Kenyon College, 201 N College Rd, Gambier, OH 43022, USA}
\affiliation{Department of Mathematics and Theoretical Physics, University of Cambridge, Wilberforce Rd, Cambridge UK, CB3 0WA}

\author{Matthew~Cerep}
\affiliation{Department of Physics and Astronomy, Swarthmore College, Swarthmore, PA 19081, USA}
\affiliation{Department of Physics and Astronomy, West Virginia University, P.O. Box 6315, Morgantown, WV 26506, USA}
\affiliation{Center for Gravitational Waves and Cosmology, West Virginia University,Chestnut Ridge Research Building, Morgantown, WV 26505, USA}

\author{Shar~Daniels}
\affiliation{Department of Physics and Astronomy, Swarthmore College, Swarthmore, PA 19081, USA}
\affiliation{Department of Physics and Astronomy, University of Delaware, Newark, Delaware 19716, USA}

\begin{abstract}

We study the nonlinear effects of minimally coupled, massless,  cosmological scalar fields on the cosmic microwave background (CMB). These fields can exhibit post-recombination parametric resonance and subsequent nonlinear evolution leading to novel contributions to the gravitational potential.  We compute the resulting contributions to the CMB temperature anisotropies through the time-variation of the gravitational potential (i.e., the integrated Sachs-Wolfe (ISW) effect). We find that fields that constitute 5\% of the total energy density and become dynamical at $z_c \simeq 10^{4}$ can produce marginally observable ISW signals at multipoles $\ell \simeq 2000$. Fields that become dynamical at earlier times and/or have initial displacements at a flatter part of their potential, produce ISW contributions that are significantly larger and at higher multipoles. We calculate these dynamics and the resulting evolution of gravitational perturbations using analytic estimates alongside detailed nonlinear lattice simulations, which couple scalar fields and cosmological fluids to a perturbed metric. Finally, we discuss the possibility of detecting these features with future high-resolution CMB observations.

\end{abstract}
\date{\today}

\maketitle
\section{Introduction}

Scalar fields are ubiquitous in cosmology, from the inflaton \cite{Guth:1980zm} that phenomenologically gives a mechanism for the flatness and structure of our observed universe, to quintessence \cite{Caldwell:1997ii} which provides a dynamical explanation for the current epoch of accelerated expansion. In between these two epochs, there have been a plethora of proposed cosmological scalar fields, with a range of masses, motivated by many beyond-the-standard model scenarios from inflationary model building \cite{Baumann:2014nda}, to string theory \cite{Svrcek:2006yi,Arvanitaki:2009fg,Kane:2015jia,Stott:2017hvl}, to attempts to explain the seemingly fine-tuned nature of the current epoch of accelerated expansion \cite{Griest:2002cu,Linder:2010wp,Kamionkowski:2014zda}. Indeed, adding scalar fields is often the first line of attack to confront inconsistencies in our cosmological scenarios.  After that, however, it is imperative that these scenarios be forced to predict how their existence will impact our cosmological observations. 

In this paper we consider the consequences of a single cosmological scalar field, which is minimally coupled, with a canonical kinetic term, and a potential that goes as $V = \lambda \varphi^4$ about its minimum. With these restrictions, the scalar field will be held fixed at some initial displacement by Hubble friction. When the Hubble parameter drops below some critical value, the field will become dynamical and eventually oscillate about the minimum of the potential. Such a field is often referred to as `Early Dark Energy' (EDE). 

There is a rich literature exploring the linear and nonlinear dynamics in cosmological scalar fields both in the very early and late universe. In the context of the post-recombination universe, a linearized analysis of the resonant growth of scalar field fluctuations, and its impact gravitational potentials was explored earlier in \cite{Amin:2011hu}. In that case, however, the scalar field was not always subdominant in energy density. A linearized analysis of resonant growth of perturbations in certain oscillating dark energy models was investigated in \cite{Johnson:2008se}. In terms of numerical simulations, the resonant growth of perturbations, and nonlinear dynamics in scalar fields for $\lambda_n \varphi^n$ potentials has been studied in detail in the  early universe context \cite{Lozanov:2016hid,Lozanov:2017hjm,Khlebnikov:1996mc}.\footnote{In general, there is a large literature on how parametric resonance and subsequent nonlinear dynamics may also play a central role in thermalizing the energy stored in the inflaton at the end of inflation, initiating the epoch of radiation domination. See \cite{Amin:2014eta,Lozanov:2019jxc} for recent reviews.} 

Indirect constraints on cosmological scalar fields which are dynamically relevant between inflation and today can be placed using observations of the cosmic microwave background (CMB) and clustering of matter (through galaxy clustering and/or weak lensing). Given that the field is minimally coupled, its effects on observables come from how it modifies the expansion history and how its perturbations gravitate. Changes to the homogeneous expansion leads to changes to Hubble friction, modifying the growth of dark matter perturbations. Given that the field oscillates about an approximately quartic minimum, non-resonant modes have significant pressure support which prevents them from growing. This, in turn, limits their impact on cosmological observables.  For fields that become dynamical during radiation domination non-resonant modes, at the linear level, act like a perfect, $w=-1/3$, fluid, with a maximum fractional energy density of $\rho_\varphi/(\rho_\varphi+\rho_\gamma + \rho_\nu) \lesssim 5\%$ \cite{Poulin:2018cxd}, where $\rho_{\gamma}$ is the energy density of the CMB photons and $\rho_\nu$ is the energy density of three standard ultra-relativistic neutrino species.

The self-interaction from the $\lambda\varphi^4$ potential allows the fluctuations in such fields to grow significantly through parametric resonance, whereby linear perturbations grow exponentially by efficiently extracting energy from the oscillating homogeneous background field. Ref.~\cite{Smith:2019ihp} noted that such a process occurs in fields which may resolve the Hubble tension. The underlying reason is that for such potentials, the resonant wave numbers do not flow out of the resonance band as the universe expands.\footnote{As discussed in Ref.~\cite{Smith:2019ihp}, this implies that the typical potential used to resolve the Hubble tension, which has $V\propto \varphi^6$ around its minimum, does not experience significant effects through parametric resonance.} The growth rate of field fluctuations eventually exceeds the Hubble rate (even if the field is subdominant in energy density), leading to nonlinear, spatially inhomogeneous evolution of the scalar field. 

In the post-recombination $\Lambda$CDM universe, and during matter domination, the gravitational potential $\Psi$ is determined primarily by the dark matter perturbations, with $\dot{\Psi}\simeq0$. The significant resonant growth of scalar field perturbations, even if its background energy density is subdominant, can lead to a significant contribution to $\dot{\Psi}$. This evolution can leave an imprint on the CMB via the Integrated Sachs-Wolfe effect (ISW). It is this effect that we explore in detail in this paper.\footnote{For fields which become dynamical well before matter-radiation equality, they may make a non-negligible contribution to gravitational potential at the resonant scale. The fact that the scalar field energy density is subdominant limits its impact.} 

Through both detailed simulations and analytic scaling equations, we show that this novel ISW contribution is larger the earlier the field becomes nonlinear. The time at which nonlinear evolution starts, in turn, is earlier if the field either becomes dynamical earlier and/or starts at a location where the potential is flatter. As a result of this we find that scalar fields which may address the Hubble tension become nonlinear too late to produce an observable ISW. For fields that become dynamical earlier/start on a flatter part of their potential, the resulting ISW can dominate over the primary CMB power spectrum, with a peak at very small angular scales which cannot be probed by current CMB measurements, but which may be accessible to proposed CMB telescopes such as CMB-HD \cite{Sehgal:2019ewc,CMB-HD:2022bsz}.

The paper is organized as follows. In Section \ref{sec:model}, we describe our fiducial model and present some details of our linear and nonlinear analyses, highlighting its important features. We also describe the metric perturbations and fluid descriptions used in our nonlinear analysis to model a realistic universe. In Section \ref{sec:linearandnonlinear} we describe our numerical methods. Section \ref{sec:res_and_decay}, we present the results of our linear analysis, showing the evolution of the homogeneous mode of EDE, as well as a comparison between our linear and nonlinear results which demonstrate the presence of parametric resonance and validate the code used for our nonlinear analysis. We then outline our calculations of the ISW from the nonlinear simulation. In Section \ref{sec:ISW} we present the resulting ISW contributions for several different scalar field models. In Section \ref{sec:discussion}, we summarize our results and discuss their significance with regards to falsifying the EDE model. Note that we work in units where $c=\hbar=k_B=1$ and where the reduced Planck mass is $\mpl \equiv (8 \pi G)^{-1/2}$.

\section{The Dynamics of A Cosmological Scalar Field} \label{sec:model}

The action for a minimally coupled scalar field, $\varphi$, with a canonical kinetic term, is given by
\Beq
    \label{eq:Lagrangian}
	S = \int d^4x\sqrt{-g} \left[ \frac{\mpl^2}{2} R - \frac{1}{2} \partial_\mu \varphi \partial^\mu \varphi - V(\varphi) +\mathcal{L}_{\lcdm}\right],
\Eeq
where $\mathcal{L}_{\lcdm}$ includes contributions from the cosmological constant, cold-dark matter and the rest of the Standard Model.  
We work in the Conformal Newtonian gauge; a perturbed, conformal time foliation of FLRW without vector and tensor perturbations,
\Beq
\label{eq:metric}
ds^{2}=a^{2}[-(1+2\Phi)d\tau^{2}+(1-2\Psi)d\mathbf{x}\cdot d\mathbf{x}],
\Eeq 
where $a$ is the scale factor.  Throughout our work here, we assume that there is no significant gravitational anisotropic stress, so that the two Bardeen potentials are the same, $\Psi = \Phi$.  The gravitational potential $\Psi$ is evaluated using Einstein's equations linearized in $\Psi$ (but not the field $\varphi$).\footnote{The consistency of these assumptions can be tested using full nonlinear gravitational methods \cite{Giblin:2018ndw}; however, we do not expect any significant corrections to our results when we relax these assumptions \cite{Giblin:2019nuv}.}
With this metric, the equation of motion for this field is
\begin{align}
\label{eq:nlKG}
\ddot{\varphi}-\left(1+4\Psi\right)\nabla^{2}\varphi+2\left(\mathcal{H}-2\dot{\Psi}\right)\dot{\varphi}&\\+a^{2}\left(1+2\Psi\right)\frac{d V}{d \varphi}&=0\nonumber
\end{align}
where an over-dot is a partial derivative with respect to conformal time, $\mathcal{H} \equiv (da/d\tau)/a$, and $\mathcal{H}$ includes contributions from both the spatially averaged scalar field, as well as the $\lcdm$ components. That is,
\Beq
\mathcal{H}^2=\frac{a^2}{3\mpl^2}\left[\bar{\rho}_\varphi+\bar{\rho}_{\lcdm}\right],\,
\label{eq:friedconst}
\Eeq
where 
\begin{equation}
\bar{\rho}_{\Lambda{\rm CDM}} = 3 H_0^2 \mpl^2\left(1-\Omega_m+\frac{\Omega_m}{a^3} + \frac{\Omega_r}{a^4}\right), \label{eq:rho}
\end{equation}
$\Omega_m$ and $\Omega_r$ is the total matter (cold dark matter plus baryons) and total radiation (photons plus neutrinos) density parameters and
a bar indicates the quantity is spatially averaged. The scalar field stress energy tensor is given by 
\begin{equation}
    (T_\varphi)^{\mu}_{\ \nu} = \nabla^\mu \varphi \nabla_\nu \varphi - \frac{1}{2} \left[\nabla^\alpha \varphi \nabla_\alpha \varphi + 2 V(\varphi)\right] \delta^\mu_{\ \nu}.\label{eq:scalar_SE}
\end{equation} 

For concreteness, we take the potential associated with the scalar field to be 
\Beq
\label{eq:pot}
V(\varphi)=m^2f^2\left(1-\cos\frac{\varphi}{f}\right)^{\!2}\,,
\Eeq
which, in the limit where $\varphi/f\ll 1$, resembles a massless scalar $V(\varphi)\simeq (m^2/f^2)\varphi^4/4$. Our choice of the precise form of $V(\varphi)$ is for convenience, and to make contact with earlier work \cite{Smith:2019ihp}. The scaling equations derived in this Section lead us to conclude that any potential which has a quartic minimum, and flattens to a constant at large field values, will have a qualitatively similar phenomenology.

Note that the effective mass near the minimum of the potential $m^2_{\rm eff}\equiv d^2V/d\varphi^2\simeq 3m^2(\varphi/f)^2$ which is different from the mass parameter $m^2$. In the following discussion equations using the `$\sim$' symbol are missing factors of order unity whereas those with `$\simeq$' retain those factors but are still approximate. 

\subsection{Background evolution}

It will prove useful to develop a set of approximate equations which summarize the background evolution of the scalar field. Taking the homogeneous limit of Eq.~(\ref{eq:nlKG}) we have
\begin{equation}
    \ddot{\bar{\varphi}} +2 \mathcal{H} \dot{\bar \varphi} + a^2 V_{,\varphi} = 0,
    \label{eq:backgroundphi}
\end{equation}
where we have introduced the notation $V_{,\varphi} \equiv dV/d\varphi  $. 
Roughly, when the conformal Hubble parameter is large enough, the friction term dominates and we have 
\begin{equation}
    \dot{\bar\varphi} \sim -\frac{a^2}{\mathcal{H}} V_{,\varphi}.
\end{equation}
During this `slow-roll' evolution the equation of state parameter associated with the field is given by 
\begin{equation}
    w_\varphi \simeq -1 + \frac{\dot{\bar \varphi}^2}{2a^2V} \sim -1 +\frac{a^2}{\mathcal{H}^2(a)} \frac{V_{,\varphi}^2}{V}.
\end{equation}
Since $w_\varphi > -1$, once the field becomes dynamical the critical scale factor, $a_c$,  is roughly given by 
\begin{equation}
    \frac{a_c^2}{\mathcal{H}^2(a_c)} \frac{V_{,\varphi}^2}{V} \sim 1.
\end{equation}
If we write $V(\bar \varphi) = m^2 f^2 E(\theta \equiv \bar \varphi/f)$, 
then this condition can be written terms of the mass parameter:
\begin{equation}
    m \sim \frac{\mathcal{H}(a_c)}{a_c} \frac{\sqrt{E}}{|E_{,\theta}|}\bigg|_{\theta = \theta_i}, \label{eq:eff_mass}
\end{equation}
where we evaluate the potential-dependent term at the initial field displacement,
$\bar{\varphi}_i=\theta_i f$.

The fractional contribution the field makes to the total energy density when it becomes dynamical can now be written 
\begin{equation}
   \frac{\bar{\rho}_\varphi(a_c)}{\bar{\rho}_{\rm tot}(a_c)} \simeq\frac{a_c^2 V(\bar{\varphi}_i)}{\mpl^2 \mathcal{H}^2(a_c)} \sim \frac{f^2}{\mpl^2} \left(\frac{E}{E_{,\theta}}\right)^2\bigg|_{\theta = \theta_i}. \label{eq:frac}
\end{equation}
After the field becomes dynamical it quickly starts to oscillate with a cycle-averaged equation of state parameter $\langle w_{\varphi} \rangle = 1/3$ with an envelope that decreases as \cite{Poulin:2018cxd}
\begin{equation}
\varphi_{\rm env}(a) = \varphi_c \frac{a_c}{a}.
\end{equation}
Note that empirically we find that $\varphi_c \simeq 0.7 \varphi_i$. 

\subsection{Linear perturbations}

If the field were a spectator during inflation then it would generically have both adiabatic and isocurvature initial conditions \cite{Hlozek:2017zzf,Smith:2019ihp}. Here we will ignore the isocurvature perturbations.\footnote{As discussed in Ref.~\cite{Smith:2019ihp}, the amplitude of the isocurvature perturbations scales with the tensor to scalar ratio, $r$. Given current upper limits on $r$ from measurements of the $B$-mode polarization, the EDE isocurvature contribution is, in general, too small to impact current measurements.} The superhorizon adiabatic initial conditions during slow-roll can be found analytically, giving \cite{Smith:2019ihp}
\begin{equation}
    \frac{\delta \varphi(a\leq a_c)}{f} \sim \frac{m^2a^2}{\mathcal{H}^2(a)} E_{,\theta}\bigg|_{\theta = \theta_i}\zeta_{\rm ad}(\vec k). \label{eq:SHevo}
\end{equation}

Outside of resonant phenomena (discussed below), linear perturbations can be characterized by the cycle-averaged non-adiabatic sound speed \cite{Poulin:2018cxd}
\begin{equation}
    \langle c_s^2 \rangle = \frac{2 a^2 \varpi^2(a)+k^2}{6 a^2 \varpi^2(a)+k^2},
\end{equation}
where $\varpi(a)$ is the angular frequency of the background field. The detailed evolution of $\varpi$ is unimportant, since at all times these perturbations will have significant pressure support, leading to $\delta \rho_\varphi/\bar{\rho}_\varphi \simeq {\rm constant}$ on subhorizon scales. 

\subsection{Parametric resonance}

\label{ap:floquet}

The fluctuations of the scalar field are unstable and undergo exponential growth in a narrow band of wavenumbers, as discussed in detail in Ref.~\cite{Smith:2019ihp} for the EDE context.  We summarize the main results here. Ignoring the effects of gravity, the Fourier modes of field perturbations at linear order satisfy
\Beq
\label{eq:phipert}
\delta\ddot{\varphi}_k+2\mathcal{H}\delta\dot{\varphi}_k+\left[k^2+a^2V''(\bar{\varphi})\right]{\delta\varphi}_k=0\,.
\Eeq
The $a^2V''(\bar{\varphi})\propto a^2\bar{\varphi}^2$ term provides a time-dependent, approximately periodic, oscillatory contribution to the effective frequency. 

Soon after the field becomes dynamical, the amplitude of the oscillations in the background field are damped so that $\varphi/f\ll 1$ and the potential can be approximated as $V \propto \varphi^4$. When the potential is well-approximated by a power-law the perturbations evolve as
\Beq
\label{eq:growth}
\delta\varphi_k(a)\simeq \delta\varphi_k(a_c)\frac{a_c}{a}\exp\left[\int_{a_c}^a \frac{\Re[\mu_k]}{\mathcal{H}} d\ln a\right].
\Eeq
 The perturbations grow exponentially fast around the resonant wavenumber $k_{\rm res}$, with a width $\Delta k_{\rm res}$, and a Floquet exponent $\mu_{k_{\rm res}}$ given by \cite{Smith:2019ihp}
\Beq
&k_{\rm res}\simeq 1.27m \left(\frac{\varphi_c}{f}\right)a_c\,,\\
&\gamma_{\rm res} \equiv \frac{\Delta k_{\rm res}}{k_{\rm res}}\simeq 0.17\,,\\
&\Re[\mu_{k_{\rm res}}]\simeq 0.036 m\left(\frac{\varphi_c}{f}\right)a_c\,.
\Eeq

The above discussion allows us to derive an approximate equation for the scale-factor when the resonant wavenumber becomes nonlinear, $a_{\rm nl}$. In the following we will assume that $a_{\rm nl} \ll 1$, ensuring that the universe is filled with just matter and radiation. Using the approximate, resonantly growing solution \eqref{eq:growth}, we can estimate when the perturbations become nonlinear:
\Beq
\label{eq:nonlinear}
\frac{\gamma_{\rm res}k_{\rm res}^3}{2 \pi^2}  P_{\dvp}(k_{\rm res},a_{\rm nl}) \equiv \gamma_{\rm res} \Delta^2_{\dvp}(k_{\rm res},a_{\rm nl}) \simeq 1,
\Eeq
where $\langle \delta \varphi_{\bm{k}} (a) \delta \varphi_{\bm{k}'}(a)\rangle=(2\pi)^3 P_{\delta \varphi/\varphi_{\rm env}}(k,a)\delta({\bm k}-\bm{k'})$. Solving for $a_{\rm nl}$ we have
\begin{eqnarray}
    \frac{a_{\rm nl}}{a_c} &\simeq& 1-\frac{14\ln [\gamma_{\rm res} \Delta^2_{\dvp}(k_{\rm res},a_{c})]}{ma_c/\mathcal{H}(a_c) \theta_c}\nonumber \\ &+& \frac{a_c}{a_c+a_{\rm eq}} \left(\frac{14\ln [\gamma_{\rm res} \Delta^2_{\dvp}(k_{\rm res},a_{c})]}{2ma_c/\mathcal{H}(a_c) \theta_c}\right)^2, \label{eq:anl}
\end{eqnarray}
where $a_{\rm eq} \equiv \Omega_r/\Omega_m$ and is the scale factor at which the matter and radiation energy densities are equal. Since fluctuations in $\varphi$ are still linear at $a_c$, both terms are positive and give $a_{\rm nl} > a_c$. 

Since $k_{\rm res} \sim m a_c \sim \mathcal{H}(a_c)$ we can approximately compute $\Delta_{\delta \varphi/\varphi_{\rm env}}(k_{\rm res},a_c)$ using the superhorizon solution given in Eq.~(\ref{eq:SHevo}) which allows us to write
\begin{eqnarray}
    \Delta_{\dvp}(k_{\rm res},a_c) &\simeq& A_s^{1/2} \frac{m^2a_c^2}{\mathcal{H}^2(a_c)}\frac{E_{,\theta}(\theta_i)}{\theta_c},\nonumber \\
    &\sim& \frac{A_s^{1/2}}{p_i},\label{eq:init_PS}
\end{eqnarray}
where $p_i \equiv d \ln E/d\ln \theta|_{\theta = \theta_i}$ is the effective power-law index of the potential at the initial field displacement\, and $\Delta^2_\zeta(k)\simeq A_s$ assuming scale invariant initial conditions.

\subsection{The ISW contribution}

\label{sec:ISWcontrib}

The ISW can be calculated using a line of sight integral \cite{Seljak:1996is} 
\begin{eqnarray}
\label{eq:ISW}
T_{\rm ISW}(\hat{n})&=&2\int_{\tau_i}^{\tau_0}d\tau \frac{\partial \Psi(\tau,\vec x)}{\partial \tau}\bigg|_{\vec x = (\tau_0-\tau)\hat{n}},
\end{eqnarray}
where $\tau_i$ is some initial conformal time and $\tau_0$ is the conformal time today.
The angular power spectrum due to the ISW effect is
\begin{eqnarray}
C_l^{\rm ISW} &=& 8 \pi \int_{\tau_i}^{\tau_0} d \tau \int_{\tau_i}^{\tau_0} d \tau'\int d\ln k \Delta^2_{\dot \Psi}(k; \tau, \tau') \\ &\times& j_l(k[\tau_0-\tau]) j_l(k[\tau_0-\tau']),\nonumber
\end{eqnarray}
where $j_l(x)$ is a spherical Bessel function. This allows us to see that since $d\tau = d \ln a/\mathcal{H}$, the ISW contribution scales as $C_l^{\rm ISW} \propto \Delta^2_{\dot \Psi/\mathcal{H}}(\ell \simeq k [\tau_0-\tau_{\rm nl}])$.

The scalar field contribution to the rate of change of the Newtonian potential on subhorizon scales ($k \gg \mathcal{H}$) can be approximated by 
\begin{equation}
   \nabla^2 \dot \Psi_\varphi \,\simeq\, \frac{a^2}{2 \mpl^2} \nabla^j T^{0}_{\ j}.
\end{equation}
where we have taken the subhorizon limit  since $k_{\rm res}$ is well within the horizon at $a_{\rm nl}$. In Fourier space, and linearizing around $\bar{\varphi}$, we have
\begin{equation}
    \frac{\dot \Psi_{\varphi}}{\mathcal{H}} \simeq \frac{1}{\mathcal{H}\mpl^2} \dot{\bar \varphi} \delta \varphi,
\end{equation}
which has a power spectrum
\begin{eqnarray}
    \Delta^2_{\dot \Psi_\varphi/\mathcal{H}}(k,a) &\simeq& \frac{\dot{\bar \varphi}^2}{\mpl^4\mathcal{H}^2}\varphi_{\rm env}^2 \Delta^2_{\dvp},\\
    &\simeq& \frac{\bar{\rho}_{\varphi}(a)}{\bar{\rho}_{\rm tot}(a)} \theta_c^2 \frac{f^2}{\mpl^2} \left(\frac{a_c}{a}\right)^2 \Delta^2_{\dvp}(k,a).\nonumber
\end{eqnarray}
The maximum contribution to the ISW effect from the scalar field will be at $k=k_{\rm res}$ and $a=a_{\rm nl}$, where $\Delta^2_{\dvp}(k_{\rm res},a_{\rm res})\simeq1$, giving
\begin{equation}
     \Delta^2_{\dot \Psi_\varphi/\mathcal{H}}(k_{\rm res},a_{\rm nl}) \simeq \frac{\bar{\rho}_{\varphi}(a_{\rm nl})}{\bar{\rho}_{\rm tot}(a_{\rm nl})}\left(\frac{f}{\mpl} \frac{a_c}{a_{\rm nl}}\theta_c\right)^2. \label{eq:ISWamp}
\end{equation}
Assuming that $a_{\rm nl} >a_{\rm eq}$,\footnote{In order to produce a measurable ISW contribution we must have $a_{\rm nl}> a_{\rm rec} > a_{\rm eq}$ so this assumption is required to produce an observable signal.} we have $\bar{\rho}_\varphi/\bar{\rho}_{\rm tot} \propto 1/a$, and using Eq.~(\ref{eq:frac}) we can write this as 
\begin{eqnarray}
    \Delta^2_{\dot \Psi_\varphi/\mathcal{H}}(k_{\rm res},a_{\rm nl}) &\sim& \left(\frac{\bar{\rho}_{\varphi}(a_c)}{\bar{\rho}_{\rm tot}(a_c)}\right)^2 p_i^2 \label{eq:ampISW2}\\ &\times&\frac{{\rm max}(a_c,a_{\rm eq})}{a_{\rm nl}}\left(\frac{a_c}{a_{\rm nl}}\right)^2.\nonumber
\end{eqnarray}
\begin{figure}[!t]
    \centering
    \includegraphics[width=\columnwidth]{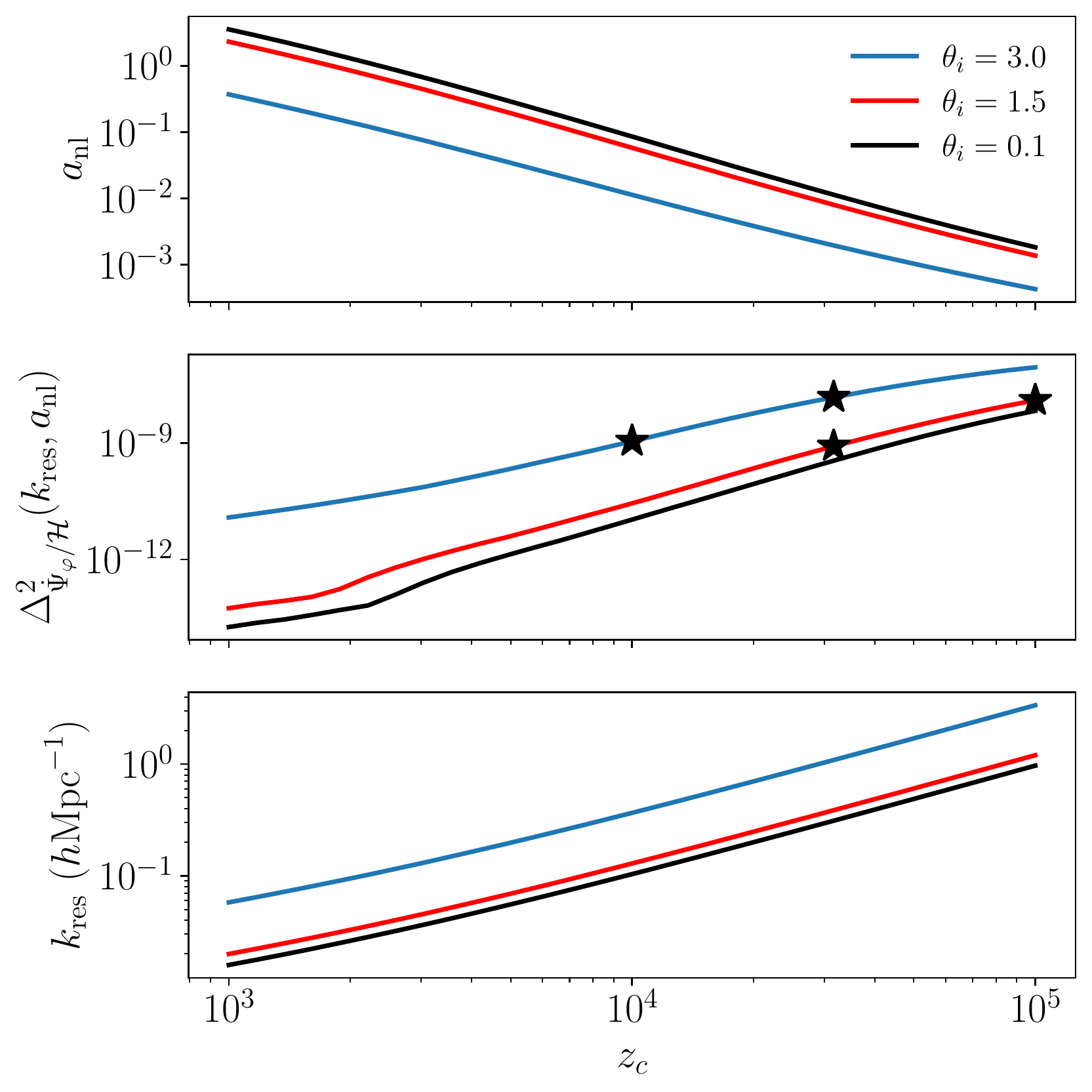}
    \caption{The scale factor when the field becomes nonlinear, $a_{\rm nl}$ (top panel), the factor that determines the amplitude of the ISW contribution [see Eq.~(\ref{eq:ISWamp})] (middle panel), and the resonant wavenumber, all as a function of the critical redshift, $z_c$, (bottom panel). The stars in the middle panel correspond the four models that we show in Fig.~\ref{fig:Hubb_spectrum}.  These locations explicitly show that decreasing $\theta_i$ decreases the ISW peak; however, a larger value of $z_c$ increases the strength of the signal. Note that when computing this figure we use Eqns.~(\ref{eq:anl}), (\ref{eq:ISWamp}, and solve for the exact background field dynamics to determine $f/m_{\rm pl}$ $m$, and $\theta_c$.}
    \label{fig:acanl}
\end{figure}

We are now in a position to anticipate what aspects of the cosmological scalar field will control its contribution to the ISW effect. First note that the analytic approximations, Eqns.~(\ref{eq:ISWamp}) and (\ref{eq:ampISW2}), indicate that the relevant dynamics depend on the shape of the potential around the initial field displacement (i.e., Eqns.~(\ref{eq:eff_mass}), (\ref{eq:frac}), (\ref{eq:init_PS})), and on the fact that the potential goes as $V \propto \varphi^4$ around its minimum. In other words, the exact shape of the potential between the initial field displacement and the minimum has a subdominant effect on the relevant field dynamics. Fixing $a_c$ and $\bar{\rho}_{\varphi}(a_c)/\bar{\rho}_{\rm tot}(a_c)$, Eq.~(\ref{eq:ampISW2}) tells us that the amplitude of the ISW contribution has a strong dependence the ratio $a_c/a_{\rm nl}$, with a smaller $a_{\rm nl} \rightarrow a_c $ leading to a larger ISW contribution. 

As we now argue, the overall ISW contribution increases as the  power-law slope of the potential at the initial field displacement decreases. First note that Eq.~(\ref{eq:ampISW2}) tells us that the ISW contribution scales as $ \propto p_i^2/a_{\rm nl}^3$. Next note that $[m\theta_c a_c/\mathcal{H}(a_c)]^{-1}\sim p_i\sqrt{E(\theta_c)/\theta_c^4}$ from Eq.~(\ref{eq:eff_mass}), so that Eq.~(\ref{eq:anl}) tells us $a_{\rm nl} \propto p_i$  or $a_{\rm nl} \propto p_i^2$, if the first or second term dominates, respectively. Therefore, the ISW contribution will roughly scale as $\propto 1/p_i^n$, with $1\lesssim n\lesssim 4$.  

Fig.~\ref{fig:acanl} shows how the different quantities that determine the amplitude of the ISW contribution depends on both the critical redshift, $z_c$, and the initial field displacement, $\theta_i$. This figure shows that for a fixed value of $\bar{\rho}_{\varphi}(a_c)/\bar{\rho}_{\rm tot}(a_c)  =0.05$, the amplitude of the ISW is set by the quantity plotted in the middle panel which shows that we can achieve a similar amplitude by increasing $z_c$ and/or increasing $\theta_i$. Note that the pairs of model parameters which have a similar amplitude in the middle panel of Fig.~\ref{fig:acanl} (indicated by the stars) also have similar ISW contributions, as shown in the right panel of Fig.~\ref{fig:Hubb_spectrum}.

\section{Linear and Nonlinear Numerical Methods}
\label{sec:linearandnonlinear}

In all cases we split the $\lcdm$ content into separate contributions from matter, radiation and dark energy.  Each of these contributions will contribute to the background evolution of the scale factor, via Eq.~(\ref{eq:rho}). Most simulations that study the parametric decay of scalar condensates contain only scalar fields and self-consistently calculate the evolution of the homogeneous spacetime.  Sometimes these simulations will either impose an expansion history, as was an option in Ref.~\cite{Felder:2000hq}, or add a homogeneously diluting component that contributes to the self-consistent evolution, e.g. Ref.~\cite{Weiner:2020sxn}. In this work we will additionally track the inhomogeneities of matter and radiation using a fluid treatment.  Since we will keep perturbation in the these fluids to linear order, we will consider these perturbations in momentum space following the method derived in Ref.~\cite{Ma:1995ey}. For each fluid, $i$, with (constant) equation of state, $w_i\equiv p/\rho$, the fluid perturbation is \Beq
\delta_i \equiv \delta \rho_i/\bar{\rho_i},
\Eeq 
which we keep throughout the simulation as a momentum-space quantity.  The variable $\theta$ is defined by
\begin{equation}
    \left(\bar{\rho}+\bar{P}\right)\theta \equiv ik^j\delta T^0_j,
\end{equation}
where the contributions to $\theta$ from each species, $\theta_i$, are just the divergences of the fluid velocities, $\theta_i = ik^j\partial_j{v_i}$, and
\begin{equation}
    \sum_i\left(\bar{\rho}_i + \bar{P}_i\right)\theta_i \equiv \left(\bar{\rho}+\bar{P}\right)\theta.
\end{equation}
We then have a set of evolution equations for the fluid variables, where $i$ can be either matter or radiation, 
\begin{align}
	\label{eq:deltadot}
	\dot{\delta}_i&=-(1+w)(\theta_i-3\dot{\Psi})\\
    \label{eq:thetadot}
    \dot{\theta}_i&=-\frac{\dot{a}}{a}(1-3w)\theta_i+\frac{w}{1+w}k^2\delta_i+k^2\Psi. 
\end{align}

Of course, both the field and fluid equations require us to know the metric perturbations.  These can be found from the linearized Einstein Equations,
\begin{align}
	\label{eq:kpoisson1}
	{\Psi} &= -\frac{1}{k^2} \left[ \tilde{S}_1 - \frac{1}{k^2} \frac{\dot{a}}{a} \left(\tilde{S_2} - \frac{3}{2 \mpl^2} a^2 \sum_{\rm species} (\bar{\rho} + \bar{P})\theta \right) \right] \\
    \label{eq:kpoisson2}
    \dot{{\Psi}} &= - \frac{1}{3k^2} \left[ -\frac{1}{k^2} \left(\tilde{S_2} - \frac{3 }{2 \mpl^2} a^2 \sum_{\rm species} (\bar{\rho} + \bar{P})\theta \right) + \frac{\dot{a}}{a} {\Psi} \right]
\end{align}
where the two quantities $\tilde{S}_1$ and $\tilde{S}_2$ are the Fourier Transforms of
\Beq
\label{eq:source1}
S_1 = \frac{1}{2 \mpl^2} \delta \rho_{\varphi} = \frac{1}{2 \mpl^2} \left(\frac{1}{2}\dot{\varphi}^2 + \frac{1}{2} \left(\nabla \varphi\right)^2 + V(\varphi) -\bar{\rho}_\varphi\right)
\Eeq
and 
\Beq
\label{eq:source2}
S_2 = \frac{3}{2 \mpl^2} \partial_i (\partial_0 \varphi \partial_i \varphi) = \frac{3}{2 \mpl^2} \left[\nabla \varphi\cdot\nabla\dot{\varphi} + \dot{\varphi}\nabla^2 \varphi \right],
\Eeq 
which are the contributions to the Poisson equations from the field.

In both the linear and nonlinear analyses we solve the Friedmann constraint, Eq.~\eqref{eq:friedconst}, alongside the fluid equations of motion Eq.~\eqref{eq:deltadot} and Eq.~\eqref{eq:thetadot} and the two Poisson equations for $\Psi$, Eq.~\eqref{eq:kpoisson1}, and $\dot{\Psi}$, Eq.~\eqref{eq:kpoisson2}.  In our linear analysis, we additionally separate out the field average from its fluctuations, $\varphi = \bar{\varphi} + \delta \varphi$, and solve for the dynamics of the homogeneous mode, $\bar{\varphi}$, via Eq.~\eqref{eq:backgroundphi} separately from the perturbations, $\delta \varphi$, see equation Eq.~\eqref{eq:phipert}. We complete this analysis entirely in momentum space.  

In our nonlinear analysis we utilized a modified version of {\sc GABE} (Grid And Bubble Evolver) \cite{Child:2012qg} that simulates the scalar field in configuration space according to Eq.~\eqref{eq:nlKG} while still evolving the fluid variables in momentum space.  The sources, \eqref{eq:source1} and \eqref{eq:source2}, are  calculated in configuration space, then Fourier transformed to allow us to invert Eqns.~\eqref{eq:kpoisson1} and \eqref{eq:kpoisson2} to find the gravitational perturbations.  These perturbations, $\Psi$ and $\dot{\Psi}$ are stored both in momentum space (to be used in the evolution equations for the fluids) and inverse Fourier transformed into configuration space (to be used in the evolution equation for the field).  This procedure allows us to solve for all linearized quantities in momentum space, while allowing us to treat the field evolution nonlinearly and  fully resolve the dynamics of the  configuration-space metric perturbations.

In order to generate initial conditions for our lattice simulations, we numerically solve the set of coupled differential equations in a simplified Einstein-Boltzmann hierarchy and approximate recombination as instantaneous. For this analysis, we include a tightly coupled baryon-photon fluid, CDM, neutrinos, and the scalar field. The neutrinos are treated as a perfect fluid (i.e. their anisotropic stress vanishes, $\sigma_\nu = 0$). We have also evolved the system with free-streaming neutrinos with a Boltzmann hierarchy that is truncated at the third moment (while still using the approximation $\Phi = \Psi$), using the proscription outlined in Ref.~\cite{Ma:1995ey}, and found no difference in the resulting field dynamics. 

In the lattice simulations we treat the CDM/baryons as a single matter fluid, and the photons/neutrinos as a single radiative constituent. We chose the following parameters for all of our simulations: $A_s=2 \times 10^{-9}$, $h = 0.67$, $\Omega_M = \Omega_c + \Omega_b = 0.314$, $\Omega_R = \Omega_\gamma + \Omega_\nu = 9.16 \times 10^{-5}$. We evolve the system from an initial scale factor, $a_i$, is set to be small enough so that the field dynamics are linear,  to $a_f=1/30$, using a time step of $\Delta t = L/N/100$, where $L$ is the length of an edge of the simulation box and $N$ is the number of pixels on one side of the box. Our fiducial lattice size is $N^3=256^3$ and we choose $L$ to ensure that the resonant wavenumber, $k_{\rm res}= 2\pi/\lambda_{\rm res}$, is well inside of the box, $L = 20 \lambda_{\rm res}$. We have confirmed that using smaller boxes do not alter our results.

\section{Resonance and nonlinear evolution of the Scalar Field}\label{sec:res_and_decay}

\subsection{The homogeneous mode}

In this Subsection we will focus on two models--where we set $\bar{\rho}_{\varphi}(a_c)/\bar{\rho}_{\rm tot}(a_c) = 0.05$, $z_c=10^{4.5}$, and choose either $\theta_i=1.5$ or $\theta_i=3$. 

We start by comparing the homogeneous evolution of the scalar field in our lattice simulation with our linear calculations.  The field $\varphi$ starts out as roughly homogeneous, and starts oscillating when $a=a_c$, where $a_c$ is determined by $a_c^2V''(\varphi(a_c))\sim \mathcal{H}^2(a_c)$. As the field enters a period of coherent oscillations, we expect that the contribution to $\bar{\rho}_{\varphi}$ ceases to look dark energy-like (approximately constant) and starts to look radiation-like -- since the minimum is massless.  For our linear analysis, the homogeneous mode will continue to oscillate about its minimum, decaying only due to Hubble friction.  However, in the nonlinear analysis, we expect the homogeneous mode to show signs of earlier decay when the field exits the linearized regime.

\begin{figure}[!t]
    \centering
    \includegraphics[width=\columnwidth]{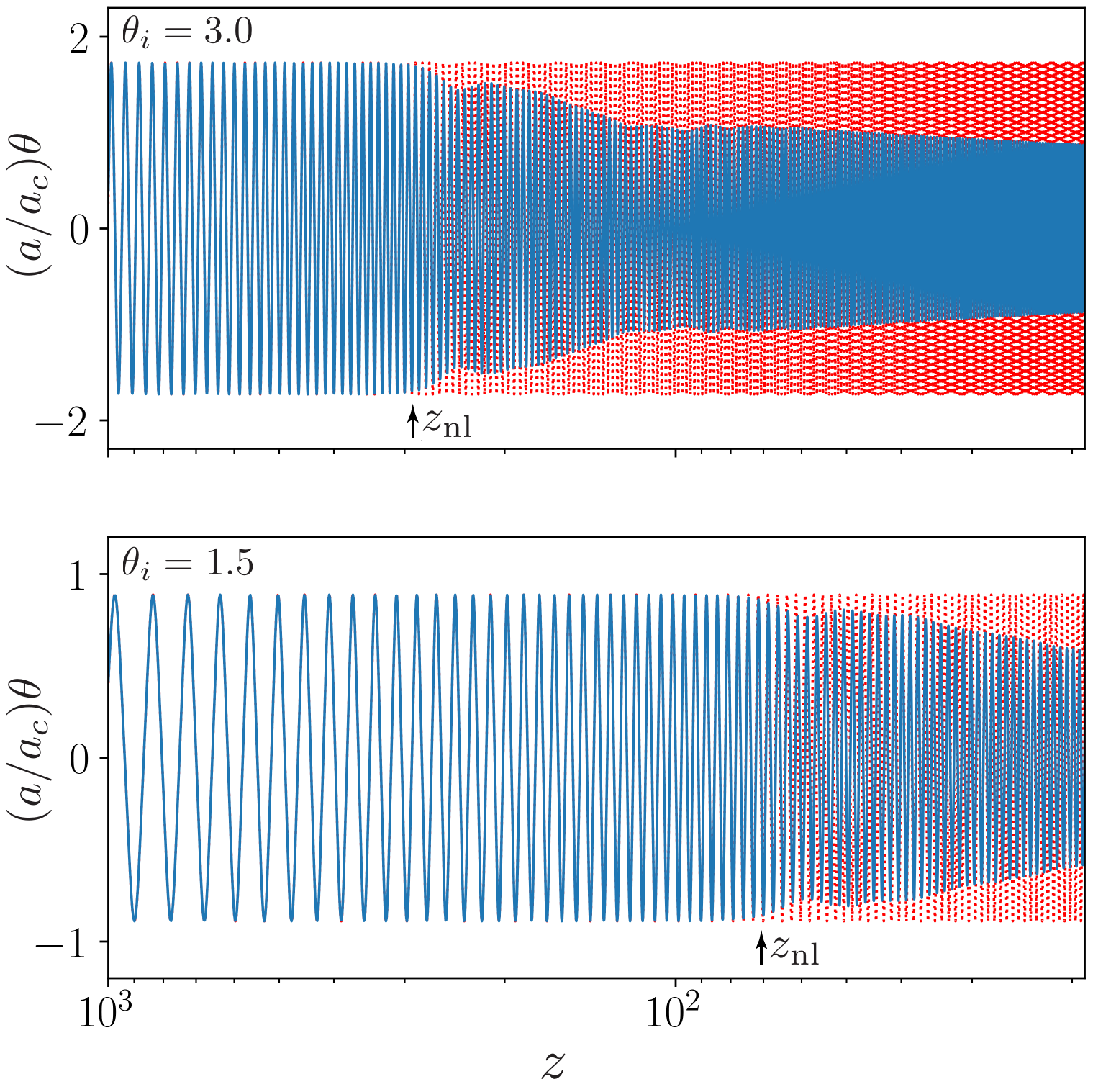}
    \caption{The background evolution of the scalar field field for $z_c=10^{4.5}$ from our nonlinear simulations (blue) compared to a background-only calculation (red) with $\theta_i = 3$ (top panel) and $\theta_i = 1.5$ (bottom panel). The decrease in amplitude in the nonlinear simulation comes from the coupling between the homogeneous mode and the perturbations and indicates that the field dynamics are nonlinear.}
    \label{fig:bkg}
\end{figure}

Fig.~\ref{fig:bkg} shows a comparison of the homogeneous evolution of the scalar field between the linear and nonliear simulations. We can see that, in each case, there exists a $z_{\rm nl}$ at which the homogenous mode in the fully nonlinear simulation starts to decay away from the linear solution.  This is due to the transfer of energy from the homogeneous mode to the $k>0$ modes, indicating the presence of resonance from the nonlinear self-coupling. 
This also gives us an opportunity to test the accuracy of the analytic equations presented in Sec.~\ref{sec:ISWcontrib}. The top panel of Fig.~\ref{fig:acanl} shows that for $z_c = 10^{4.5}$, $\theta_i=3.0$ has $z_{\rm nl} \simeq 500$ and for $\theta_i = 1.5$ has $z_{\rm nl} \simeq 120$. Comparing this to the values of $z_{\rm nl}$ denoted in Fig.~\ref{fig:bkg} ($z_{\rm nl} = 300$ and $z_{\rm nl} = 70$, respectively) shows that the analytic formulae provide a good approximation to the redshift of nonlinearity within a factor of order unity. 

\begin{figure}[!t]
    \centering
    \includegraphics[width=\columnwidth]{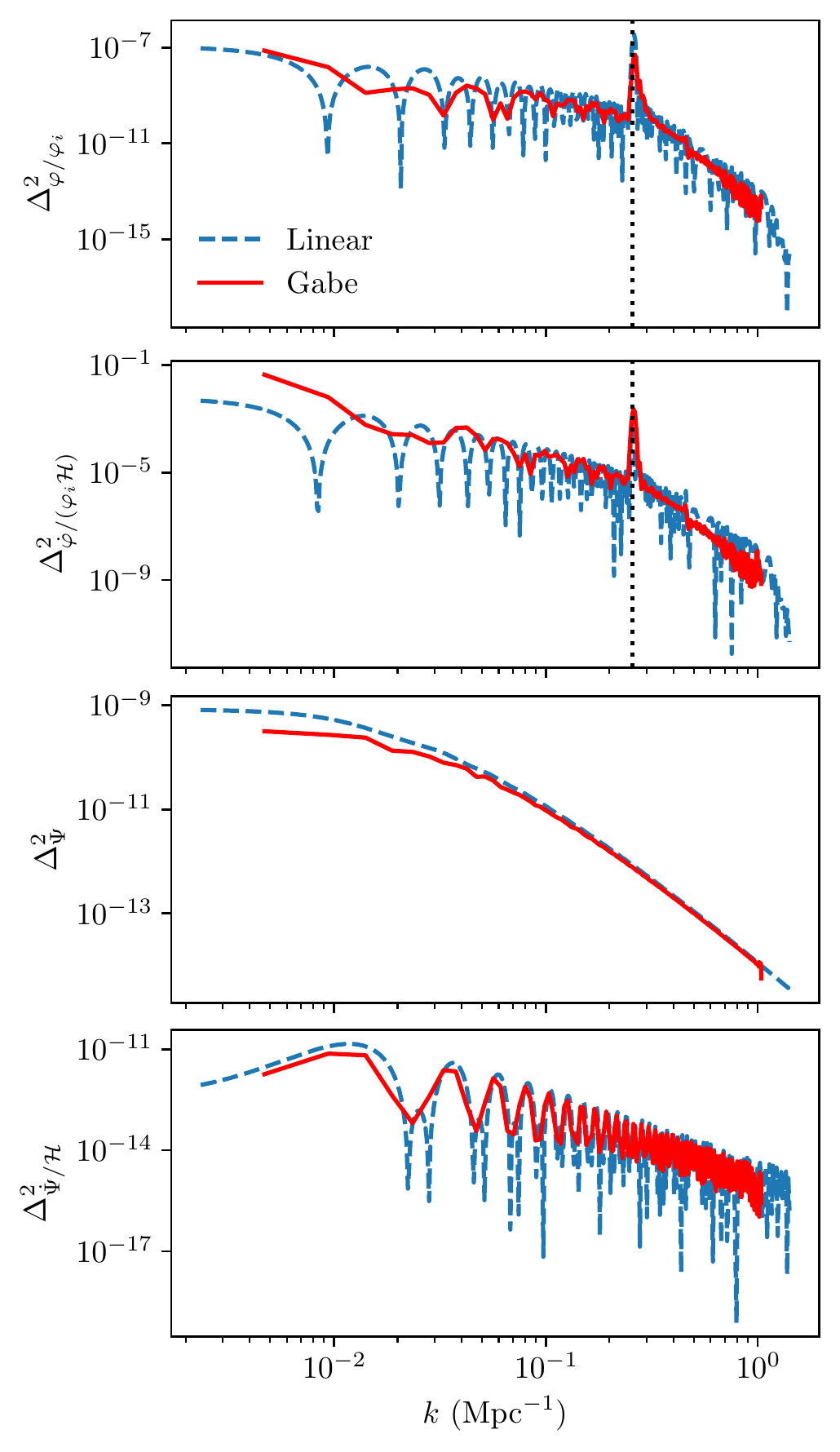}
    \caption{A comparison between the dimensionless power spectra for different quantities at a later time in the simulation (in this case $z_c = 10^{4.5}$ and $\theta_i=1.5$).  For each panel, the blue dashed line shows the  power spectrum from a linear analysis at $z=500$, while the red (solid) curves come from the $z=500$ slice from a run initialized at $z=1000$. The resonant wave number, $k_{\rm res} \simeq 0.26\ {\rm Mpc}^{-1}$, is indicated by the dotted line in the top two panels.}
    \label{fig:comp_spectra}
\end{figure}

\subsection{The inhomogeneous dynamics}

We begin by validating our simulations by showing consistency between our linear and nonlinear analyses at a time when the nonlinear simulations have significantly evolved, but where we still expect to see excellent agreement with the linear analysis.  Fig.~\ref{fig:comp_spectra} shows how well the nonlinear simulations reproduce the mode-by-mode behavior of the radiation and matter fluids, the field and the gravitational perturbations when compared to the linear simulations. It is clear that both the field and fluid dynamics are being solved correctly at the linear level.  Small differences between the two spectra at large scales are due to sample variance and binning in the nonlinear simulations. Note that the Newtonian potential, $\Psi$, is mainly sourced by the matter, and its time derivative is mainly sourced by the (oscillating) radiation content.  

\begin{figure}[!t]
    \centering
    \includegraphics[width=\columnwidth]{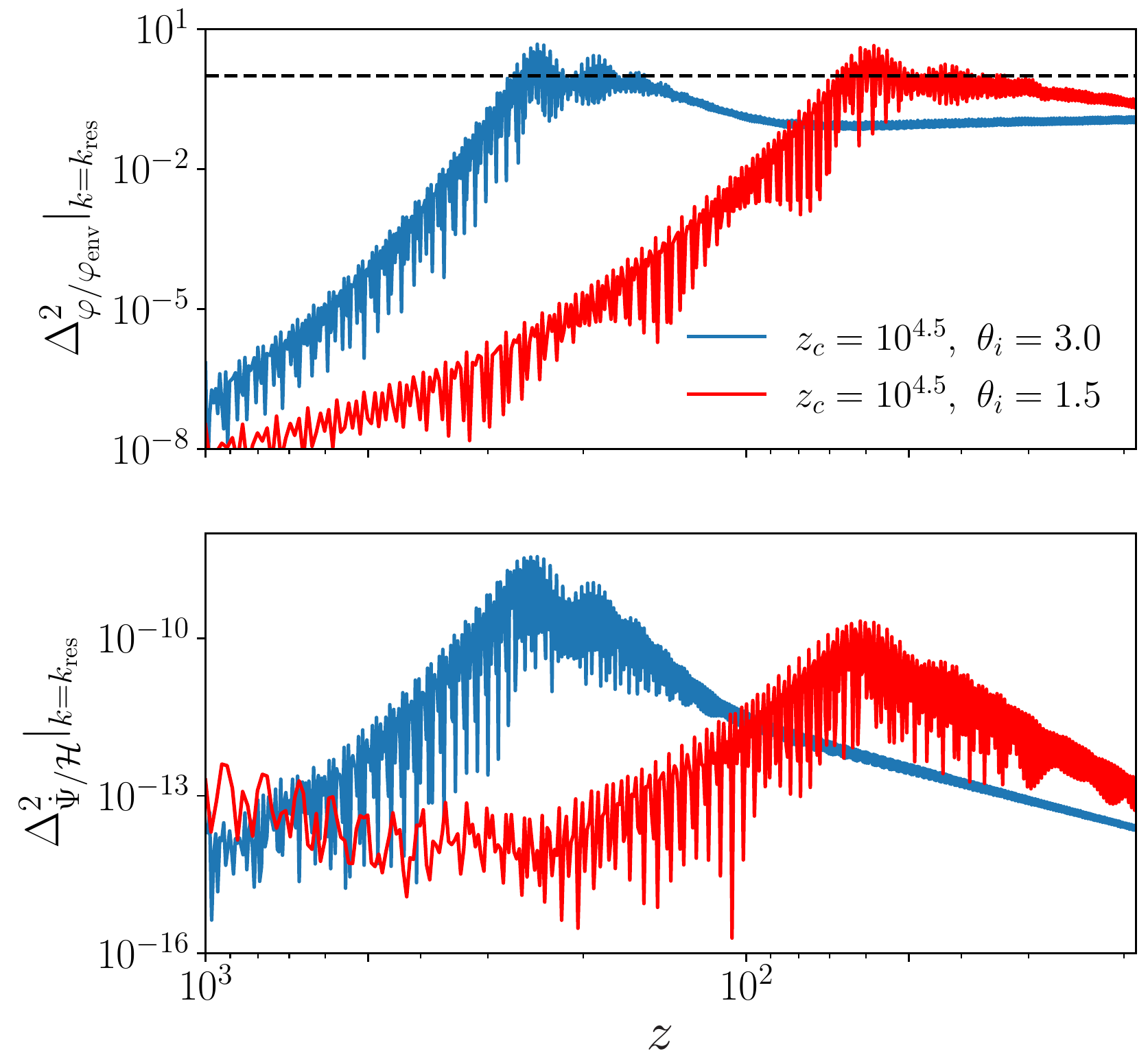}
    \caption{The evolution of the scalar field dimensionless power spectrum normalized by the field's envelope at the resonant wavenumber (top panel). We can see that once the field becomes nonlinear the field's dimensionless power spectrum is relatively constant. The dimensionless power spectrum of $\mathcal{H}^{-1} \partial \Psi/\partial \tau$ at the resonant wavenumber. As discussed in Sec.~\ref{sec:ISWcontrib}, this quantity measures the contribution that the scalar field makes to the late-time ISW effect (bottom panel).  In both panels the simulation with $\theta_i = 3$ is shown in blue and $\theta_i = 1.5$ in red. }
    \label{fig:P_kres}
\end{figure}

In order to further make a connection between the full simulation and the analytic expressions in Sec.~\ref{sec:ISWcontrib}, we plot the evolution of the resonant wavenumber in Fig.~\ref{fig:P_kres}. As expected, once the field becomes nonlinear the perturbations remain relatively constant. The resulting contribution to the ISW is shown in the bottom panel of Fig.~\ref{fig:P_kres}. There we can see that this contribution peaks at $z_{\rm nl}$ and, as expected from the middle panel of Fig.~\ref{fig:acanl}, the peak with $\theta_i = 3$ is higher than $\theta_i=1.5$. We can make this comparison more quantitative by noting that Eq.~(\ref{eq:ampISW2}) gives $\Delta^2_{\dot \Psi/\mathcal{H}}|_{k=k_{\rm res}; \theta_i=3} = 1.5 \times 10^{-8}$ and $\Delta^2_{\dot \Psi/\mathcal{H}}|_{k=k_{\rm res}; \theta_i=1.5} = 8 \times 10^{-10}$ which is within a factor of a few of the results from the simulation in the bottom panel of Fig.~\ref{fig:P_kres}, which give $3.5 \times 10^{-9}$ and $2 \times 10^{-10}$, respectively. 

\subsection{Calculating the ISW contribution}

\begin{figure}[ht!]
    \centering
    \includegraphics[width=\columnwidth]{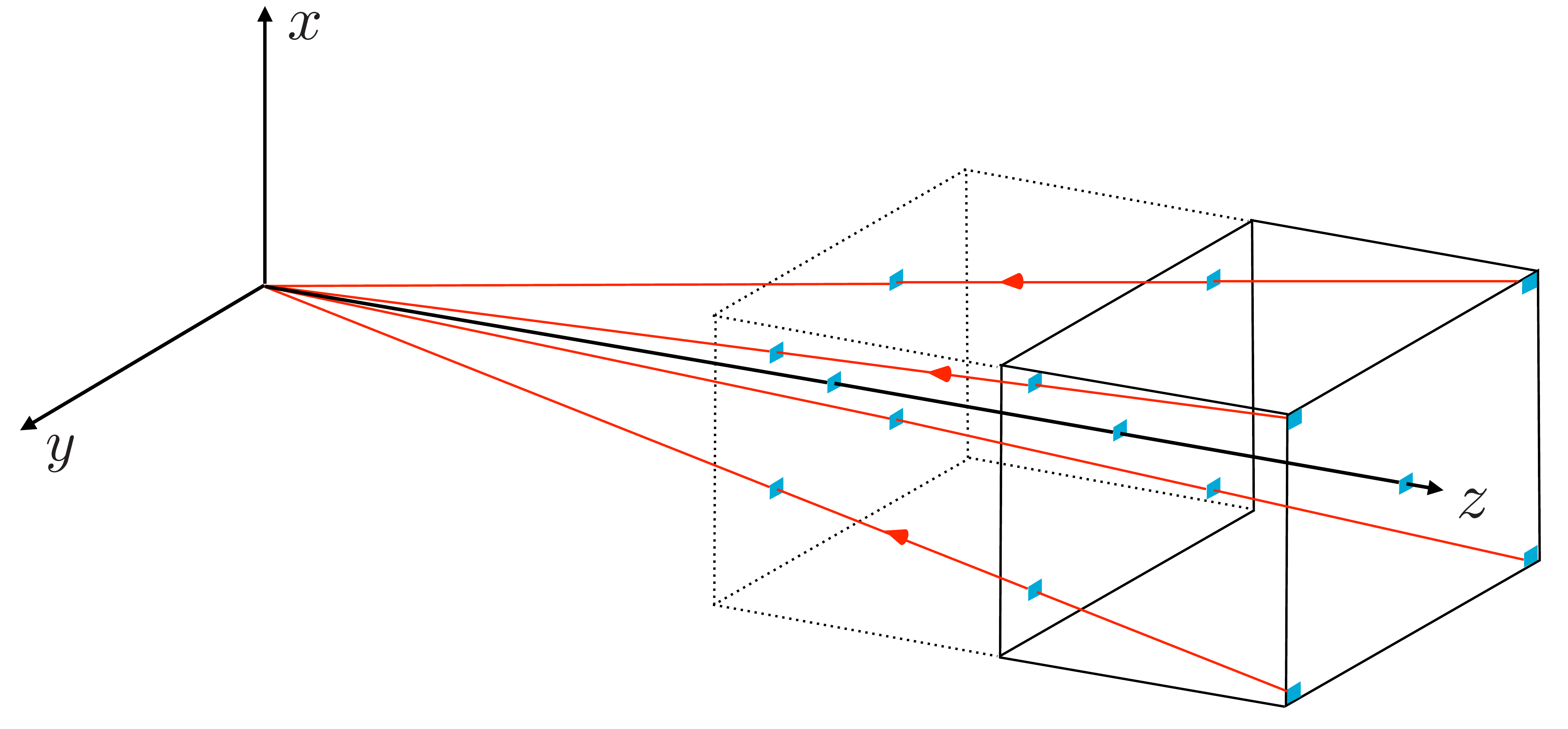}
    \caption{A cartoon describing the line-of-sight integration used to compute the scalar field's ISW contribution, given in Eq.~(\ref{eq:ISW}).}
    \label{fig:LOS}
\end{figure}

The contribution to the ISW is calculated by computing Eq.~(\ref{eq:ISW}) along several lines of sight through the simulation box; a cartoon of the procedure is given in Fig.~\ref{fig:LOS}. Slices are taken along the line of sight through the box, from $x=0$ to $x=L$. We use periodic boundary conditions, and so after $x=L$ our next slice is taken at $x=0$, and we integrate over the box again. We compute this integral from some initial redshift $z_i \gg z_{\rm nl}$ up to a time that is late enough, $z_f$, so that the ISW contribution at the final redshift slice is negligible (as shown in Fig.~\ref{fig:P_kres} the scalar field's contribution to $\dot \Psi$ decreases rapidly after $z_{\rm nl}$). We note that repeatedly traversing the same box will introduce some spurious correlations. However, given that the light-crossing time is larger than the dynamical time-scales for $k_{\rm res}\gg 1/L$, we expect they will be small for modes well within the simulation box.

\begin{figure*}[!t]
    \centering
    \includegraphics[width=\columnwidth]{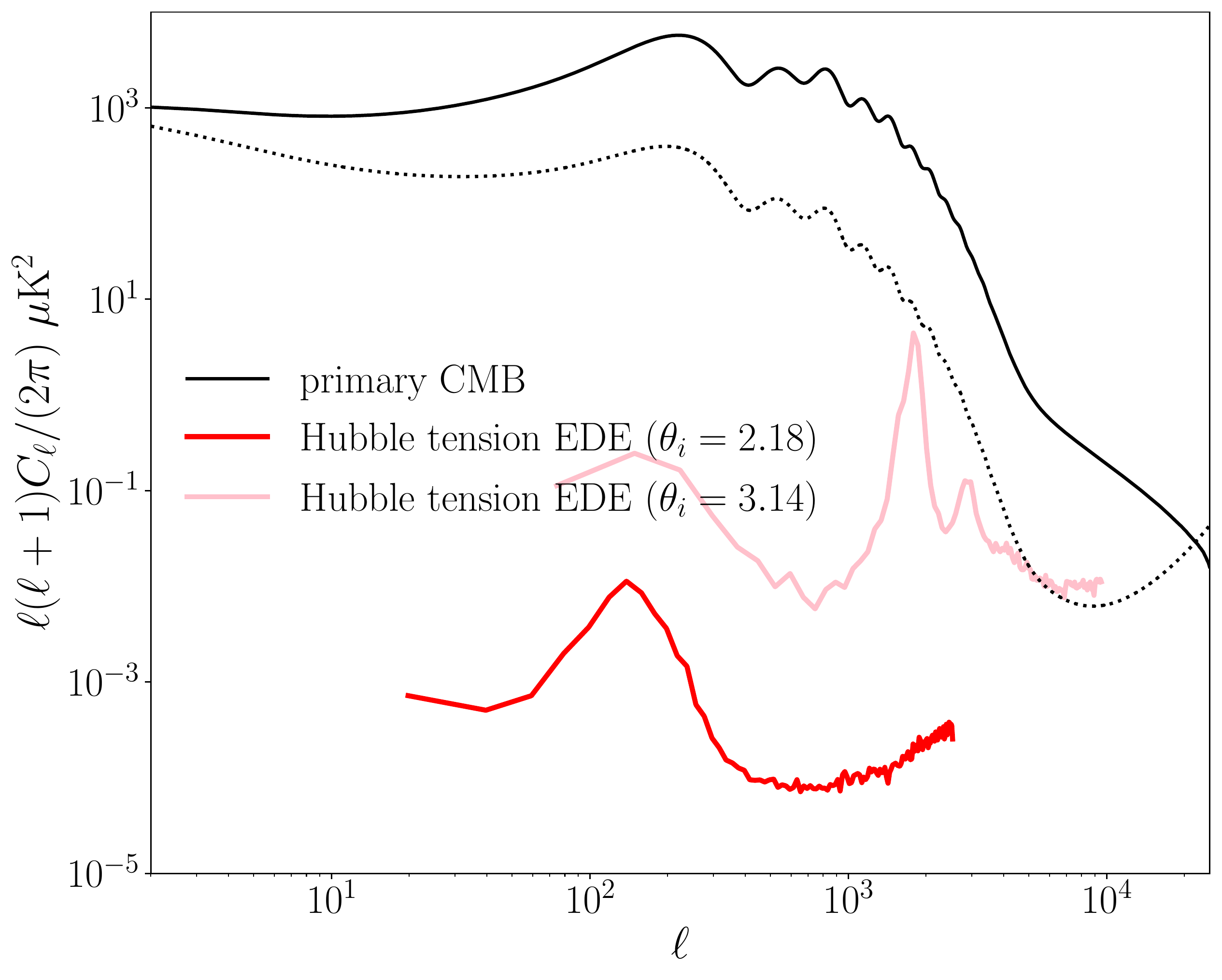}
    \includegraphics[width=.49\textwidth]{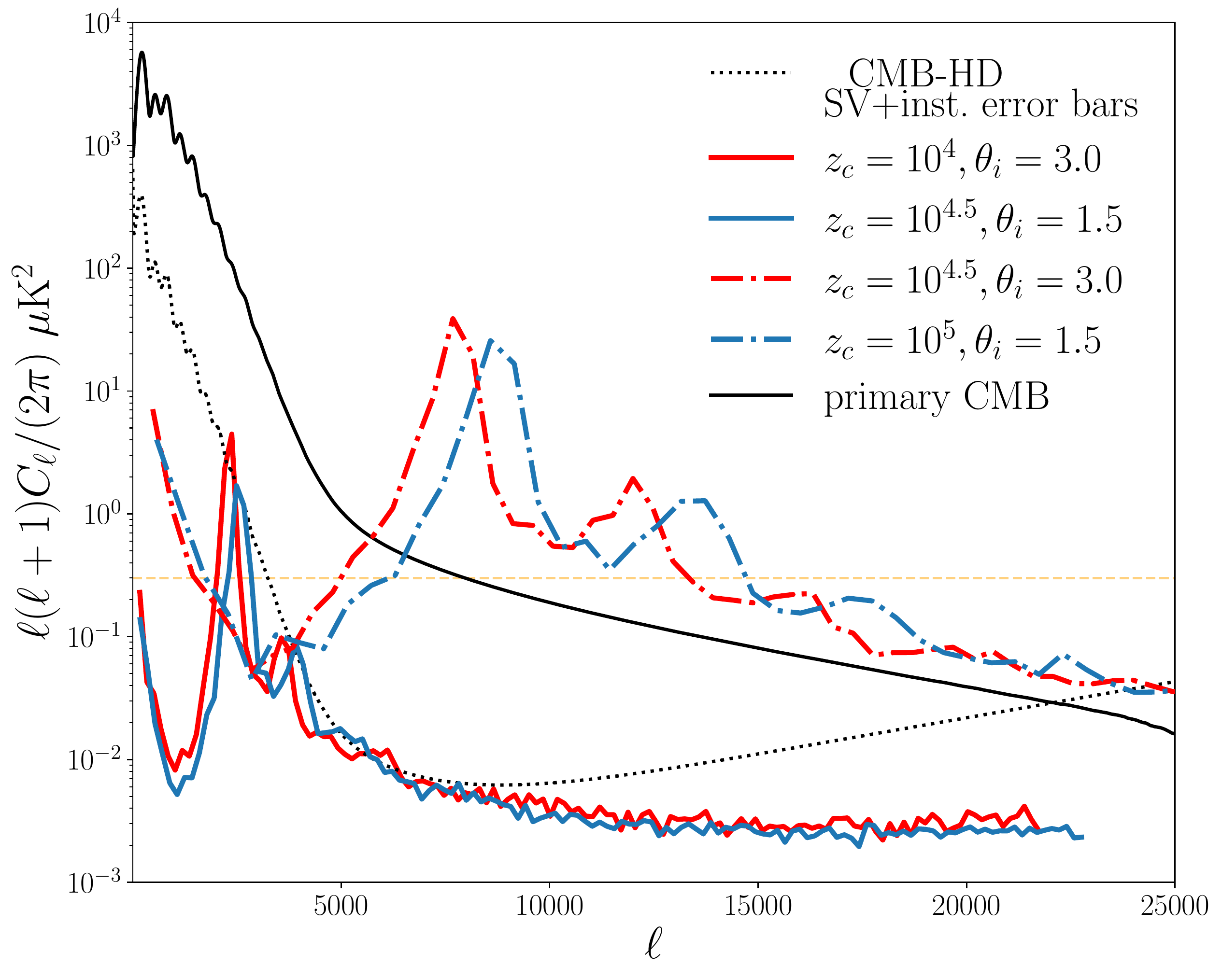}
    \caption{The ISW contribution from a scalar field that addresses the Hubble tension with $\theta_i=2.18$ (red) and $\theta_i=3.14$ (pink) compared to the primary CMB (black) (left panel). The ISW contribution from four different cosmological scalar field models (right panel). Unlike the left panel, the chosen parameters are not constrained to resolve the Hubble tension. The dashed orange curve in the right panel shows the residual foreground contribution required to achieve the science goals of CMB-HD.}
    \label{fig:Hubb_spectrum}
\end{figure*}

Once we have evaluated \eqref{eq:ISW}, we end up with a partial-sky map of the contribution to the ISW from $\dot{\Psi}$ from our nonlinear simulations.  We convert the spatial location of each ray at $z_f$ to an angular position on our sky and compute the resulting 2D Fourier transform of the angular map to determine the ISW contribution to the CMB temperature power spectrum. We have confirmed that the power spectra we compute is insensitive to moving the final redshift slice closer to today. 

\section{The scalar field ISW contribution}
\label{sec:ISW}

We are now in a position to compute the ISW contribution for this cosmological scalar field. In order to convert the temperature power spectra to units of $\mu{\rm K}^2$, we take $T_{\rm CMB} = 2.7255\ {\rm K}$ \cite{2009ApJ...707..916F}. 

\subsection{The contribution from a scalar field that address the Hubble tension}

It is of interest to determine whether a scalar field which may address the Hubble tension may include nonlinear dynamics which will produce a significant contribution to the late-time ISW effect. The best-fit parameters for such a field is given in the Appendix of Ref.~\cite{Smith:2019ihp}: $\log_{10} z_c = 3.52$, $\rho_{\varphi}(a_c)/\rho_{\rm tot}(a_c) = 0.09$, $\theta_i = 2.18$ which corresponds to $m=6 \times 10^{-28}\ {\rm eV}$ and $f/\mpl = 0.2$. These parameters lead to $k_{\rm res} = 0.043\ {\rm Mpc}^{-1}$ and $z_{\rm nl} \simeq 3.3$, which imply that the ISW contribution will peak at $l_{\rm peak} \simeq k_{\rm res}(\tau_0 -\tau_{\rm nl}) \simeq 300$. 

As shown in Eq.~(\ref{eq:ampISW2}), the overall contribution to the ISW effect is primarily determined by the ratio $a_c/a_{\rm nl}$, and in the case of a field that resolves the Hubble tension this factor is $\sim 10^{-3}$, which already gives us an indication that the ISW from this model will be very small. Indeed, as shown in the left panel of Fig.~\ref{fig:Hubb_spectrum}, we can see that the resulting power spectrum is at most $\sim 10^{-2}\ (\mu{\rm K})^2$, well below the cosmic variance limit. Also note, as we anticipated, the contribution shows a peak at $l \sim 100$. 

The main reason that this model makes such a small ISW contribution is the wide gap between $a_c$ and $a_{\rm nl}$.  Since $a_c \sim a_{\rm eq}$ in this case, the ISW contribution scales as $(a_c/a_{\rm nl})^3$ leading to a suppression of order $10^{-9}$. Fig.~\ref{fig:acanl} shows that, with $z_c$ fixed in order to resolve the Hubble tension, the only way to move these two redshifts closer is by increasing $\theta_i$ (and thereby move towards a flatter part of the potential). The pink curve in the left panel of Fig.~\ref{fig:Hubb_spectrum} shows how the ISW contribution grows in amplitude and moves to a smaller scale as $\theta_i \rightarrow \pi$. In this limit we also have $V''(\varphi)<0$, raising the possibility of a tachyonic instability. However, as shown in Ref.~\cite{Smith:2019ihp}, this instability is only present for isocurvature perturbations, and even in that case, leads to a relatively small enhancement of the perturbations. Note that, for this specific form of the scalar field potential, in order to resolve the Hubble tension we must have $\theta_i \lesssim 2.5$ at the 95\% confidence level \cite{Smith:2019ihp}. However, other forms for the scalar field potential (such as $\alpha$-attractors \cite{Braglia:2020bym}) may both resolve the Hubble tension and allow the field to start at a flatter part of its potential. 
 
\subsection{The contribution from a general scalar field}

If we allow the scalar field parameters to vary we can produce a measurable ISW effect. As discussed in the previous Section, the main way to boost the ISW contribution is to move $z_{\rm nl}$ closer to $z_c$. The top panel of Fig.~\ref{fig:acanl} shows that the most effective way to do this is to increase $z_c$. The bottom pane of Fig.~\ref{fig:acanl} shows that by increasing $z_c$ we will also cause the peak of the ISW contribution to shift to smaller scales, i.e., larger multipoles. 

Fig.~\ref{fig:Hubb_spectrum} (right panel) shows the ISW generated by four possible cosmological scalar fields. The parameters were chosen so as to produce contributions with similar amplitudes by exchanging a lower value of $z_c$ for a larger value of $\theta_i$ (note that these four models are marked in the middle panel of Fig.~\ref{fig:acanl} by the four stars). All four models have $\bar \rho_{\varphi}(a_c)/\bar \rho_{\rm tot}(a_c) = 0.05$ (saturating the 95\% upper limit from the linear effects of these scalar fields on current cosmological data). 

The power spectra show two clear peaks, which correspond to the `fundamental' and to the `first harmonic' of the resonant wavenumber. At smaller scales the nonlinear scalar field ISW effect produces a scale-invariant tail. The rise of the power spectra at the lowest multipoles is, at least in part, due to edge effects in the simulation, and thus the behavior at these multipoles should not be taken as a physical result of our analysis. 

Although current CMB measurements do not have the angular resolution and sensitivity to detect these features, future observatories will. The dotted line in Fig.~\ref{fig:Hubb_spectrum} shows the error bars due to sample variance and instrumental noise associated with CMB-HD \cite{Sehgal:2019ewc,CMB-HD:2022bsz}, a proposed high-resolution CMB ground-based observatory which would image half of the sky. Note that the noise curve does not include contributions from residual uncertainty from the subtraction of foreground contributions. CMB-HD will need to remove foregrounds to a level that ensures the instrument noise is comparable to or higher than the residual foreground contribution for $\ell >7,500$ in order to achieve its CMB lensing science goals \cite{Han:2021vtm}, corresponding to a foreground level of about $\simeq 0.3\ \mu{\rm K}^2$.

\section{Conclusions}
\label{sec:discussion}

In this paper we have calculated the ISW effect from subdominant massless scalar fields, which can potentially play a role in addressing the Hubble tension, in the late universe. We present both analytic estimates as well as high-resolution lattice simulations.  We not only included the full nonlinear evolution of the scalar field, but also included the evolution of (linearized) radiation and matter fluid perturbations, as well as metric perturbations sourced by the inhomogeneous scalar field and the fluids.  From these nonlinear simulations, we have evaluated the ISW contribution to photons traveling through the simulation, validated our results against CLASS in the appropriate limit, and compared these results, for a set of test cases, to the sensitivity of CMB-HD, a proposed CMB observatory which is designed to make resolution measurements.

As anticipated by our scaling equations, the full nonlinear simulations show that the amplitude of the ISW contribution is mainly determined by the ratio $a_c/a_{\rm nl}$. This is shown through a dramatic increase in the strength of the ISW effect when choosing parameters that ensure nonlinearity sets in at an earlier time.  The impact can also be enhanced by assuming larger initial field displacements, which lead to stronger, earlier resonance.  

While there is wide theoretical motivation for studying extra scalar degrees of freedom, a recent and popular invocation of such a model has been to ease the emerging Hubble tension \cite{Poulin:2023lkg} via an EDE field.  From what we show here, early dark energy with a potential of the form $V = m^2 f^2 (1-\cos \varphi/f)^2$, that can resolve the Hubble tension will produce a contribution to the ISW that is unlikely to be constrained or validated in upcoming CMB experiments.  We make this choice as a concrete proof-of-concept, although there has been some indication that such a potential can be realized from high energy physics \cite{McDonough:2022pku,Cicoli:2023qri}.  However, models which use a different form of the potential-- such as $\alpha$-attractors \cite{Kallosh:2013hoa,Kallosh:2013yoa,Galante:2014ifa,Braglia:2020bym}-- may resolve the Hubble tension just as well and at an initial field value where the potential is flatter producing a larger ISW effect. The flattening of the potentials in such models can also bring in additional non-perturbative dynamical effects \cite{Lozanov:2017hjm} at early times. We leave an exploration of different scalar field potentials to future work. 

Although we have established that for some model parameters the EDE ISW is larger than the noise in an instrument like CMB-HD (see Fig.~\ref{fig:Hubb_spectrum}), it remains to be seen whether this signal can be distinguished from expected foreground contamination at these small angular scales. We note that the specific harmonic structure of the EDE ISW contribution, relative to the smooth foreground power spectra, may aide in its detectability. 

In addition to the effect on the evolution of the gravitational potentials, the nonlinear fragmentation of the scalar field also source gravitational waves at frequencies of order $k_{\rm res}/a_{\rm nl}$. The post recombination production of gravitational waves at such small frequencies might lead to additional $B$-mode polarization at reionization (e.g., Ref.~\cite{Geller:2021obo}). It may also be possible to see the ISW contribution using other probes (i.e., 21cm observations \cite{Raccanelli:2015lca}), and would be worth considering in the future.

Cosmological scalar fields provide a rich phenomonology which touches all aspects of cosmology, from inflation to the current epoch of accelerated expansion, from linear to resonant nonlinear dynamics. Here we have shown that scalar fields may make a novel contribution to the ISW effect, imparting characteristic features in the very small-scale CMB temperature anisotropies. This adds further motivation to build a CMB observatory targeting the very small-scale anisotropies (such as CMB-HD) which will give us access to new ways to understand the origin and evolution of the universe. 

\acknowledgments

We thank Kaloian Lozanov for significant initial involvement, including setting up preliminary scalar-field  lattice simulations and calculations of the corresponding ISW contribution. We thank Neelima Sehgal, Dongwong Han, and Amanda Macinnis, for discussions about CMB-HD. We thank Yacine Ali-Ha\"imoud, Daniel Grin and Zachary Weiner for prompt, detailed and insightful feedback on the manuscript. The numerical work presented here utilized the Strelka Computing Cluster, which is run by Swarthmore College, and the Vera Rubin Cluster at Kenyon College, which is supported by the National Science Foundation (NSF) and the Kenyon College Department of Physics.
T.L.S.~is supported by the NSF, Grant No.~AST-2009377 and the Research Corporation. J.T.G., M.G., and E.F.~are supported by the NSF, Grant No. PHY-2013718. 
M.G.~is supported by the U.S. Department of Energy, Office of Science, Office of Advanced Scientific Computing Research, Department of Energy Computational Science Graduate Fellowship under Award Number DE-SC0023112. M.~A.~is supported by a NASA ATP-Theory Grant 80NSSC20K0518.

This report was prepared as an account of work sponsored by an agency of the United States Government. Neither the United States Government nor any agency thereof, nor any of their employees, makes any warranty, express or implied, or assumes any legal liability or responsibility for the accuracy, completeness, or usefulness of any information, apparatus, product, or process disclosed, or represents that its use would not infringe privately owned rights. Reference herein to any specific commercial product, process, or service by trade name, trademark, manufacturer, or otherwise does not necessarily constitute or imply its endorsement, recommendation, or favoring by the United States Government or any agency thereof. The views and opinions of authors expressed herein do not necessarily state or reflect those of the United States Government or any agency thereof.

\bibliography{bibliography}

\newpage

\end{document}